# Breaking the VE-cadherin bonds

Julie Gavard[1,2]

1. Institut Cochin, Universite Paris Descartes, CNRS (UMR 8104), Paris, France.
2. Inserm, U567, Paris, France.

*Corresponding author:* Julie Gavard, Institut Cochin, Cell Biology Department, Bldg Mechain, 3rd Floor, Rm. 306, 22 rue Mechain, 75014 Paris.

*Abstract:* Exchanges between the blood compartment and the surrounding tissues require a tight regulation by the endothelial barrier. Recent reports inferred that VE-cadherin, an endothelial specific cell-cell adhesion molecule, plays a pivotal role in the formation, maturation and remodeling of the vascular wall. Indeed, a growing number of permeability inducing factors (PIFs) elicits signaling mechanisms culminating at VE-cadherin destabilization and global alteration of the junctional architecture. Conversely, anti-PIFs militate against VE-cadherin disruption and enhance cell cohesion. These findings shed a new light on how endothelial cell-cell junctions impact the vascular network, and change our perception about normal and aberrant angiogenesis.

Under physiologic conditions and depending on the vascular sites, the exchange of fluids, cells, and nutrients between the blood compartment and the surrounding tissues can be increased or decreased. Endothelial cells that line the vascular wall act thus as gatekeepers to control the infiltration of blood proteins and circulating cells to the underlying tissues. This vascular permeability contributes to normal angiogenesis, blood pressure control, as well as immune responses [1]. Moreover, many pathological conditions and human diseases exhibit an abnormal increase in vascular permeability, such as tumor-induced angiogenesis, inflammation, macular degeneration, allergy, and brain stroke. It has been described that endothelial permeability can be mediated by transcellular and paracellular pathways, where plasma molecules and cells can pass through (transcellular) or between (paracellular) endothelial cells. Transcellular passage requires either cell fenestration or a complex system of transport vesicles, called vesiculo-vacuolar organelles. The paracellular pathway, by contrast, is mediated by the coordinated opening and closure of endothelial cell-cell junctions. This latter function must therefore be tightly regulated to maintain the endothelial integrity.

**1. Endothelial cell-cell junctions required functional VE-cadherin adhesion**

The adhesive contacts between cells underlie many morphogenetic processes during embryonic development as well as growth control, turnover, and regeneration of adult tissues. The formation, maintenance, and remodeling of the intercellular contacts implies a functional interaction between two main adhesive structures: tight junctions and adherens junctions (Figure 1). Cadherins, the main constituent of the adherens junctions, belong to a conserved family of adhesion molecules that provide a molecular bond between cells and link the plasma membrane with the intracellular actin cytoskeleton through catenin family proteins [2]. Tight junctions form a dense ultra-structure organization, observable by electron microscopy, which involves numerous adhesive molecules, including the claudin family of tetraspan transmembrane proteins, occludins, and junctional adhesion molecules (JAMs), as well as the intracellular adapters, namely zona occludens proteins, ZO-1 and ZO-2 [3]. Whereas in epithelial cells, tight junctions are often located apically with respect to adherens junctions, in endothelial cells, both junctions are intermingled throughout cell-cell contact areas [4].

The barrier properties require the adhesive activities of VE-cadherin and claudin-5, which are key components of the adherens and tight endothelial junctions, respectively (Figure 1). Indeed, macromolecule passage is enhanced through claudin-5 and VE-cadherin-deficient endothelial monolayers *in vitro* [5,6]. In endothelial cells, VE-cadherin is highly expressed and located at adherens junctions. Its expression profile appears early and restricted into the endothelial cell lineage, such that its promoter has been used to trigger transgene expression specifically in the endothelial compartment [7]. In addition, N-cadherin is also found at high levels in endothelial cells but exhibits a weak clustering within endothelial cell-cell junctions and occupied preferentially apical locations. Indeed, N-cadherin was suggested to play a role in vascular stabilization through heterotypic adhesion between endothelial cells and pericytes [8]. Interestingly, while claudin-5 knockout mice develop normally but have a defective blood brain barrier function and die shortly after birth [9], VE-cadherin knockout mice are embryonic lethal and exhibit multiple severe defects during developmental angiogenesis [10]. In agreement, VE-cadherin is required to prevent the disassembly of the blood vessel walls [11], and to coordinate the passage of macromolecules through the endothelium *in vitro* [5,12]. Finally, VE-cadherin has been recently shown to directly enhance the expression level of claudin-5 by tethering repressive transcription factors away from the claudin-5 promoter [6]. Conversely, the absence of functional VE-cadherin is associated with loss of claudin-5 expression, suggesting that VE-cadherin adhesion might act upstream in the formation and maintenance of the endothelial barrier integrity. In conclusion, VE-cadherin function in endothelial cells is intimately linked to vascular integrity and endothelial barrier plasticity.

**2. Molecular basis for VE-cadherin adhesive function**

VE-cadherin belongs to the cadherin super-family of cell-cell adhesion molecules, which constituted more than two hundred genes encoding cadherins in the human genome [13]. Among them, classical cadherins are $Ca^{2+}$-dependent, homophilic, cell to cell adhesion molecules expressed in nearly all cells within solid tissues. These molecules also participate in cell-cell recognition, a property that confers the ability of cells to aggregate with and ultimately sort their most physiologically relevant cell partners. It is generally admitted that cadherins form a core adhesion complex that consists of a cadherin dimer, binding through its extracellular region to another dimer of cadherins expressed in adjacent cells, and an intracellular region, anchored to the plasma membrane and linked to the cytoskeleton [14].

Similarly to other typical cadherins, the VE-cadherin extracellular domain consists of five cadherin-type repeats, called EC (extracellular cadherin) domains that are bound together by calcium ions in a rod-like structure [14]. Interestingly, VE-cadherin lacks the HAV tripeptide, localized in the EC1 repeat, which has been discussed to participate in cell-cell adhesion recognition in the case of E- and N-cadherin-mediated contacts. Based on information obtained by the use of purified recombinant proteins containing either fragments or total ECs, it is observed that multiple EC interactions promote the formation of a completely inter-digitated hexameric configuration. Of note, this hexameric structure might be specific for VE-cadherin, and has been proposed to favor its membrane stability [15]. Once engaged in homophilic interactions, the extracellular domains develop additional forces, strengthening the adhesive contacts while retaining the ability to associate and to dissociate rapidly in response to dynamic changes in the cellular environment.

Classical cadherins are single-pass transmembrane proteins that interact with a number of different cytoplasmic partners to carry out their functions, which include cell-cell adhesion, cytoskeletal anchoring, immediate intracellular signaling and transcriptional control. The cadherin-associated proteins, catenins, are universally present in classical cadherin complexes [13]. In particular, β-catenin interacts with the distal part of the cadherin cytoplasmic domain and p120 catenin with a more proximal region. α-catenin does not bind directly to cadherin, but instead associates with β-catenin, and provides a physical links to the actin cytoskeleton, either by directly binding to actin filaments or indirectly through other actin-binding proteins such as vinculin, α-actinin, and formins thus regulating actin polymerization, cross-linking and dynamics at the cell-cell contact zones [16]. In addition, β-catenin could be substituted to γ-catenin, which has been suggested to associate preferentially with vimentin network, and therefore may provide a link between VE-cadherin-mediated junctions and the intermediate filament cytoskeleton [4]. On the other hand, p120 catenin seems to influence cadherin function by a variety of mechanisms, which includes plasma membrane stability, recycling and cell surface targeting [17].

Therefore, disrupting the endothelial cell-cell junctions implies to disconnect VE-cadherin intracellular domain from essential cohesive mediators, as well as weakening its extracellular homophilic interactions.

### 3. How do permeability inducing factors (PIFs) disrupt VE-cadherin bonds?

There is a growing number of angiogenic and inflammatory agents that have been shown to modulate vascular integrity *in vivo* and barrier properties of endothelial monolayers *in vitro*. They can be collectively named as permeability inducing factors (PIFs), independently of their mode of actions (Table 1). Several studies have primarily focused on their effects on the endothelial cell-cell junction organization and have permitted to unveil key molecular and cellular mechanisms involved in the endothelial barrier remodeling. Four interconnected pathways can be elicited by PIFs to increase endothelial permeability, leading ultimately to the loss of VE-cadherin function, through: 1) phosphorylation-driven VE-cadherin/catenin complex destabilization, 2) cell surface expression, 3) crosstalk to tight junctions, and 4) tension and mechanical forces (Figure 2).

First, phosphorylation has been described early as a prominent mechanism by which cadherins and catenin interaction can be modulated, either positively or negatively, and therefore might decipher for overall adhesion forces. To this regard, VE-cadherin bears nine putative phospho-tyrosine sites, among which Y658, Y685 and Y731 have been individually found to be implicated in the barrier integrity [18,19]. In addition, S665 has been identified as a pivotal regulator of VE-cadherin exposure at the plasma membrane [5]. Of note, the vascular endothelial growth factor (VEGF) was demonstrated to trigger VE-cadherin phosphorylation and remodeling in endothelial cells [20]. Similarly, most of the PIFs tested so far are associated with an increase in VE-cadherin tyrosine phosphorylation and diminution of VE-cadherin/catenin binding. While several kinases have been reported to contribute to VE-cadherin phosphorylation status, the exact molecular mechanisms by which such tyrosine phosphorylation promotes vascular leakage are still elusive [4]. However, the non-receptor tyrosine kinases of the Src family appear involved in this biological effect. Indeed, Src-deficient mice show decreased vascular leakage and VE-cadherin tyrosine phosphorylation notably in response to VEGF and histamine [21]. More striking, blocking Src activity can efficiently restore the barrier integrity even in pathological context such as brain stroke, tumor cell extrasavation and retinal hyper-permeability [22-24].

Second, reduction of VE-cadherin stability at the plasma membrane might contribute to compromise endothelial barrier integrity. Interestingly, internalization of VE-cadherin has been observed in response to monocyte extrasavation, inhibition of FGF signaling and activation of VEGF-R2 by VEGF [5,25,26]. Indeed, VEGF triggers the hierarchical activation of the Src tyrosine kinase, Vav2, a guanine-nucleotide exchange factor for the small Rho GTPases, Rac and its downstream effector PAK, the p21-activated kinase [5]. This signaling axis culminates at the PAK-dependent phosphorylation of VE-cadherin, on a highly conserved serine residue, which directs VE-cadherin to endocytosis [5]. Of note, active forms of PAK have been reported to be localized at cell-cell junctions upon PIF exposure [27]. Alternatively, decreasing the association between VE-cadherin and p120 catenin leads to clathrin-dependent VE-cadherin endocytosis [28]. Indeed, blocking FGF signaling in endothelial cells triggers the dissociation between VE-cadherin and p120, leading to disassembly of adherens junctions and impairment of vascular integrity [26]. Finally, VE-cadherin cell surface expression is frequently altered in hyper-permeability conditions such as inflammation, diabetes, and virus exposure [29-31]. Ultimately, PIF dose, kinetics and mode of action might control the fate of internalized VE-cadherin and drive it to temporary storage compartment, recycling pathway, or degradation.

Third, increase in endothelial monolayer permeability is generally accompanied by the reorganization of junctional proteins and a subsequent increase in paracellular permeability. Not only VE-cadherin has been found to be relocated and phosphorylated upon diverse PIF stimulation, but similar observations have also been made for tight junction components [32]. Interestingly, the diminution of VE-cadherin plasma membrane exposure corroborates with compromised organization of the endothelial tight junctions [5,26]. As reported in epithelial cells, these observations suggest a functional crosstalk between VE-cadherin and tight junctions [3]. This hypothesis has been further demonstrated in endothelial cells lacking VE-cadherin expression or its adhesive activity. Indeed, in such cells, claudin-5 mRNA is down regulated, while rescuing VE-cadherin function triggers claudin-5 transcription [6]. Hence, VE-cadherin can directly control claudin-5 expression, and therefore the organization and maturation of endothelial tight junctions. At the molecular level, VE-cadherin adhesion prevents from β-catenin and FoxO binding to claudin-5 promoter where they could repress its transcription. These data offer a fine molecular mechanism by which VE-cadherin encrypts overall endothelial cell-cell junction architecture. On the other hand, the loss of junctional adhesion molecule, JAM-C, expression results in stabilization of VE-cadherin-mediated adhesion, while its expression corroborates with an increase in the endothelial permeability [33]. Despite these recent advances, little is still known on the crosstalk between adherens and tight junction in endothelial cells.

Finally, endothelial permeability largely depends on actomyosin-based cell contractility, as intracellular stress fibers exert centripetal tension on intercellular junctions [34]. Inhibition of acto-myosin contractility by pharmacological agents, as well as examination of myosin light chain (MLC) phosphorylation status had placed mechanical forces in the pathway by which PIFs, such as thrombin and histamine exert their effects [35]. MLC phosphorylation mainly relies on the activation of the small GTPase, RhoA, which in turn controls two serine/threonine kinases, namely ROCK and PRK [35-37]. Interestingly, elevation of intracellular calcium signaling has been shown upstream RhoA activation in such biochemical routes [37,38]. In addition, blocking PAK-dependent acto-myosin contractility prevents from endothelial permeability increased by several PIFs, such as VEGF, histamine, TNFα (tumor necrosis factor), LPS (lipopolysaccharide from the bacterial wall), as well as atherosclerotic risk factors [27,39].

In conclusion, it is most likely that coordinated disruption of VE-cadherin intracellular interactions by phosphorylation, internalization, and mechanical forces contributes to the destabilization and the disengagement of VE-cadherin adhesion, culminating at the subsequent opening of endothelial cell-cell junctions.

**4. Looking for anti-permeability inducing factor (anti-PIFs) mechanisms**

Few angiogenic mediators have been identified based on their anti-permeability action, although blocking vascular leakage may have direct implications in modulating angiogenesis and inflammation, as well as therapeutic potentials in the treatment of many human diseases characterized by loss of vascular integrity. In this paragraph, I will discuss the proposed mechanisms on the endothelial barrier function of anti-permeability inducing factors (anti-PIFs), among them: angiopoietin-1 and its cognate receptor Tie2, FGF (for fibroblast growth factor), Robo-4, cAMP-elevating G protein-coupled receptor (GPCR) agonists and S1P (sphingosine-1-phosphate) (Table 2 and Figure 2).

First, angiopoietin-1, such as VEGF, is a potent pro-angiogenic factor, but whereas VEGF causes vascular permeability, angiopoietin-1 stabilizes blood vessels and protects from VEGF-induced vascular permeability. Indeed, angiopoietin-1 administration or overexpression in the dermal compartment can protect from the potentially lethal actions of VEGF as a consequence of uncontrolled plasma leakage [40]. In this regard, angiopoietin-1 can potently block VEGF-induced endothelial permeability *in vitro* [41], suggesting that their opposing effects on vascular leakage may be exerted through direct stimulation of endothelial cells. In addition, angiopoietin-1 might exert a general anti-vascular permeability effect, protecting blood vessels from the plasma leakage caused by thrombin and bacterial wall components, such as LPS [41,42]. Thus, the angiopoietin-1/Tie2 endothelial signaling axis might play a key anti-inflammatory role in various diseases such as asthma, rheumatoid conditions and septic shock. Several signaling mediators, including calcium signaling or GTPase Activating Proteins (GAPs) for RhoA, have been proposed to oppose the endothelial barrier disruption [42,43], but the exact molecular mechanisms were still unknown. Based on our recent findings, we investigated whether angiopoietin-1 might impede on the phospho-serine-dependent internalization of VE-cadherin [5]. Indeed, we have obtained evidence that angiopoietin-1 elicits a signaling pathway through Tie2, which can compete for Src activation by VEGF-R2, therefore halting the VEGF signaling to VE-cadherin internalization [44]. Similar Src inhibition has been demonstrated as well to oppose VEGF-induced vascular permeability in response to Robo-4, a ligand for the endothelial Slit-2 receptor [45]. Interestingly, Tie2 is localized at cell-cell junctions in an intact endothelial monolayer, where it contributes to stabilize VE-cadherin adhesion, in association with the vascular endothelial phospho-tyrosine phosphatase (VE-PTP) [46,47]. Both Tie2 localization and downstream signaling are modified in the absence of endothelial cell-cell contacts, suggesting a regulatory feedback between VE-cadherin adhesion and Tie2 function. This "super" clustering of adhesion and signaling receptors at the contacting zones might then preserve controlled paracellular permeability.

In addition to angiopoietin-1 action, membrane stability of VE-cadherin is also involved in the protective mechanism exerted by FGF signaling during vascular maturation [26]. Indeed, inhibition of FGF signaling reduces the interaction between p120 and VE-cadherin, while VE-cadherin internalization is enhanced [26]. Interestingly, such interaction has been shown to be critical to prevent from VE-cadherin endocytosis [28]. Finally, cAMP-Epac-Rap1 signaling promotes decreased paracellular permeability by enhancing VE-cadherin-mediated adhesion, in response to cAMP-elevating G protein-coupled receptor (GPCR) agonists, such as prostaglandin E2 and atrial natriuretic peptide (ANP) [12,48,49]. Cytoskeletal rearrangement and barrier enhancement through Rac activation have been on the other hand proposed to militate for vascular integrity by S1P signaling [50], while reinforcement of endothelial cell/pericyte interaction through N-cadherin adhesion cannot be excluded in its anti-PIF effects [8]. Discovery of novel anti-PIFs has accelerated in the past years and research in this field had direct implications in the search for therapeutic drugs designed to target aberrant vascular leakiness, inflammation, and edema.

Thus, recent reports suggest that protection of VE-cadherin adhesion largely contributed to anti-PIF molecular mechanisms, such as through angiopoietin-1, FGF, and intracellular Rap signaling, while actin rearrangement and cell adhesion collectively can also control the endothelial barrier properties.

Despite our progresses in the understanding of the molecular mechanisms regulating VE-cadherin function in the endothelial barrier, the dynamics of VE-cadherin trafficking, including endocytosis and recycling are not elucidated yet. Ultimately, the biochemical route by which VEGF, angiogenic factors and oncogenes modulate

VE-cadherin, cell-cell junctions and vascular integrity may help identify new therapeutic targets for the treatment of many human diseases that exhibit aberrant vascular leakage. Of note, past efforts in angiogenesis field had led to the approval of anti-VEGF drugs in colon cancer treatment and ocular diseases. However, this drug is not suitable for all patients, can affect normal vasculature and exhibit tumor recurrence upon therapy withdraw. Hence, it will be crucial in the future to ascertain the molecular basis for the development of novel therapeutic targets designed to promote normalization of the vascular wall and its micro-environment.

**Acknowledgements:**
The author truly regrets that she could not cite many seminal works owing to space limitations. The author would like to thank N. Bidère (Inserm, Villejuif, France) for helpful discussions and comments. The author is supported by funding from the Centre National de la Recherche Scientifique (CNRS) and the CNRS "Projets Exploratoires/Premier Soutien" (PEPS) program.

**Competing interest statement:**
The author declares no competing financial interests.

Table 1. Permeability inducing factors (PIFs) and their proposed modes of action

| PIFs | Effects on VE-cadherin | Proposed Signaling pathways |
|---|---|---|
| VEGF | Phosphorylation<br>Internalization<br>Catenin dissociation | Src<br>Rac/PAK<br>$Ca^{++}$/PKC |
| Thrombin | Catenin dissociation<br>Phosphorylation | $Ca^{++}$/PKC<br>Rho<br>Rac/PAK<br>MAPK |
| Histamine | Phosphorylation<br>Catenin dissociation<br>Internalization | $Ca^{++}$/PKC<br>Rho<br>Rac/PAK |
| TNFα | Phosphorylation | $Ca^{++}$/PKC<br>Rho<br>Rac/PAK |
| LPS | Phosphorylation | Rho<br>Indirect<br>Src |
| ROS | Phosphorylation<br>Catenin dissociation | Rac/PAK<br>Src<br>$Ca^{++}$/PKC |

*LPS: LipoPolySaccharide; PAK; p21-activated kinase; PKC: protein kinase C; ROS: reactive oxygen species; TNF: Tumor Necrosis Factor; VEGF: Vascular Endothelial Growth Factor. Some PIFs are not listed here, such as bradykinine, as their effects have been suggested to be specific of the endothelial cell models used.*

Table 2. Anti-Permeability inducing factors (anti-PIFs) and their effects on the endothelial barrier

| Anti-PIFs | Signaling system | Effects on the endothelial barrier |
|---|---|---|
| Angiopoietin-1 | Tie-2<br>Rho-GAP<br>mDia | VE-cadherin membrane stabilization<br>Src - $Ca^{++}$ pathway inhibition<br>Acto-myosin contractility |
| Robo-4 | Slit-2 | VE-cadherin membrane stabilization<br>Src inhibition |
| S1P | S1P-R1 | Rac/PAK |
| cAMP-GPCR | Rap | VE-cadherin membrane stabilization |
| FGF | FGF-R1 | VE-cadherin membrane stabilization |

*cAMP-GPCR: cyclic adenosine monophosphate elevating G-protein coupled receptor; FGF: fibroblast growth factor; GAP: GTPase Activating Protein; mDia: mammalian diaphanous; PAK: p21-activated kinase; R: receptor; S1P: sphingosine 1 Phosphate.*

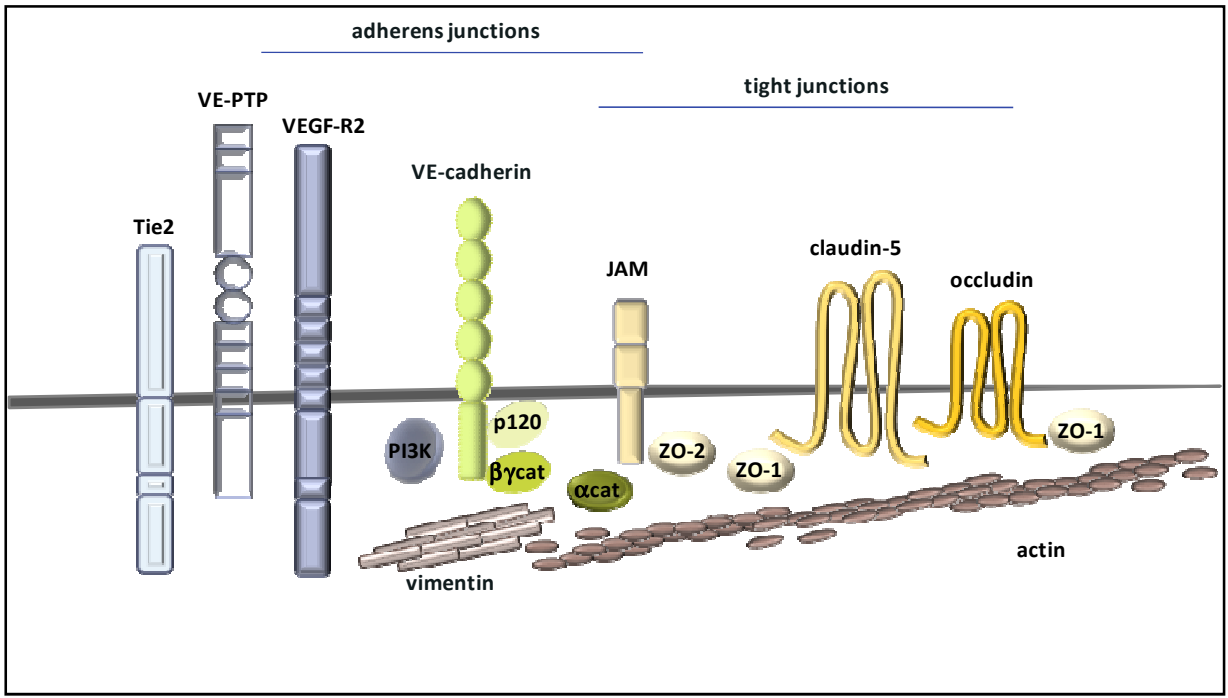

**Figure 1. Endothelial cell-cell junctions.**
Adherens and tight junctions sealed the endothelial cell contacts through specific cell-cell adhesion molecules: VE-cadherin, JAM (junctional adhesion molecules), claudin-5, occludin. Through intracellular mediators: p120 catenin, α, β and γ catenins (cat), zona occludens (ZO) proteins, they are linked to the intracellular cytoskeleton of actin and vimentin. VE-cadherin has also been characterized for its interaction with signaling component such as: phospho-inositide 3 Kinase (PI3K), vascular endothelial growth factor receptor type 2 (VEGF-R2), vascular endothelial cell-specific phospho-tyrosine phosphatase (VE-PTP) and the angiopoietin-1 receptor, Tie2.

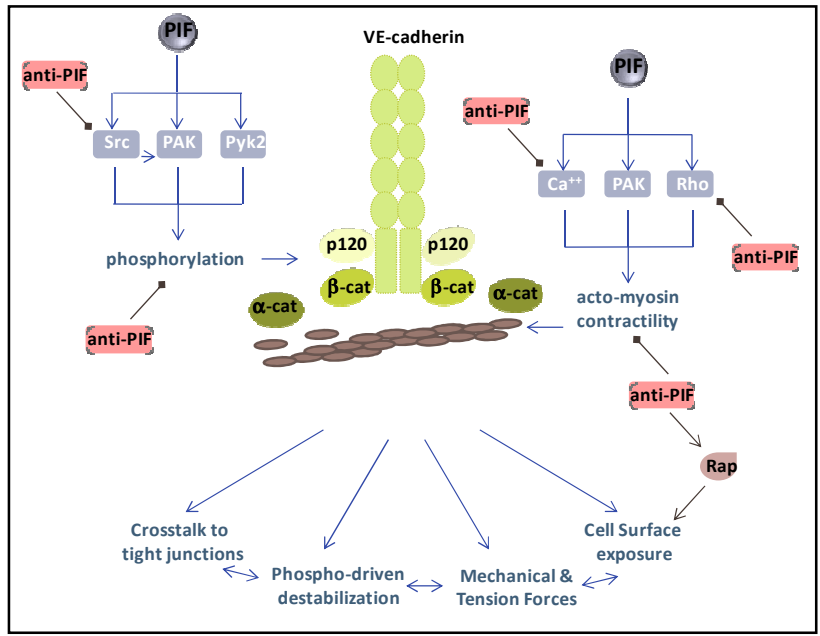

**Figure 2. VE-cadherin adhesion controls the endothelial barrier properties.**
Four interconnected pathways can be elicited by permeability inducing factors (PIFs) to increase endothelial permeability, leading ultimately to the loss of VE-cadherin function, through: 1) mechanical and tension forces, 2) phosphorylation-driven VE-cadherin/catenin complex destabilization, 2) cell surface expression, 3) crosstalk to tight junctions. Essential mediators such as the kinases: Pyk2, PAK and Src and the biochemical routes involving the RhoA and Calcium signaling axis. At the opposite, anti-PIFs elicit signaling mechanisms hampering on Src and Rho activation, calcium signaling, VE-cadherin phosphorylation, acto-myosin contractility and actin polymerization, or enhancing VE-cadherin cell surface exposure through Rap signaling axis. Ultimately, anti-PIFs can reinforce VE-cadherin adhesion and therefore the integrity of the endothelial barrier.